\documentclass[preprint]{aastex}
\usepackage{color}
\usepackage{amsmath}
\usepackage{graphicx}
\usepackage{natbib}
\usepackage[bookmarksnumbered=true,
            bookmarksopen=false,
            colorlinks=true,
            pdfborder={0 0 1},
            citecolor=blue,
            linkcolor=red,
            anchorcolor=blue,
            urlcolor=blue,
            breaklinks=true
            ]{hyperref}
\usepackage{breakurl}
\usepackage{multirow}

\begin{document}

\title{The Formation and Early Evolution of a Coronal Mass Ejection and its Associated Shock Wave on 2014 January 8}
\author{Linfeng Wan\altaffilmark{1,2}, Xin Cheng\altaffilmark{1,2}, Tong Shi\altaffilmark{1,2}, Wei Su\altaffilmark{1,2,3}, M. D. Ding\altaffilmark{1,2}}
\affil{\altaffilmark{1}School of Astronomy and Space Science, Nanjing University, Nanjing 210023, China}
\affil{\altaffilmark{2}Key Laboratory for Modern Astronomy and Astrophysics (Nanjing University), Ministry of Education, Nanjing 210023, China}
\affil{\altaffilmark{3}Key Laboratory of Dark Matter and Space Astronomy, Purple Mountain Observatory, CAS, Nanjing 210008, China}
\email{xincheng@nju.edu.cn}

\begin{abstract}
In this paper, we study the formation and early evolution of a limb coronal mass ejection (CME) and its associated shock wave that occurred on 2014 January 8. The extreme ultraviolet (EUV) images provided by the Atmospheric Imaging Assembly (AIA) on board \textit{Solar Dynamics Observatory} disclose that the CME first appears as a bubble-like structure. Subsequently, its expansion forms the CME and causes a quasi-circular EUV wave. Interestingly, both the CME and the wave front are clearly visible at all of the AIA EUV passbands. Through a detailed kinematical analysis, it is found that the expansion of the CME undergoes two phases: a first phase with a strong but transient lateral over-expansion followed by a second phase with a self-similar expansion. The temporal evolution of the expansion velocity coincides very well with the variation of the 25--50 keV hard X-ray flux of the associated flare, which indicates that magnetic reconnection most likely plays an important role in driving the expansion. Moreover, we find that, when the velocity of the CME reaches $\sim$600 km s$^{-1}$, the EUV wave starts to evolve into a shock wave, which is evidenced by the appearance of a type II radio burst. The shock's formation height is estimated to be $\sim$0.2$R_{sun}$, which is much lower than the height derived previously. Finally, we also study the thermal properties of the CME and the EUV wave. We find that the plasma in the CME leading front and the wave front has a temperature of $\sim$2 MK, while that in the CME core region and the flare region has a much higher temperature of $\ge$8 MK.
\end{abstract}

\section{INTRODUCTION}
Coronal mass ejections (CMEs) are the most eruptive phenomena in our solar system and are able to release a vast amount of magnetized plasma into the interplanetary space \citep{1985JGR....90.8173H}, thus being a potential risk for space weather near the Earth \citep{1991JGR....96.7831G,1993JGR....9818937G}. In the past decades, white-light coronagraph observations have revealed that many CMEs have three structural components: a bright loop-like front, a dark cavity underneath, and an embedded bright core \citep{1983JGR....8810210I}. The three components correspond to the region of compressed plasma, the helical magnetic flux rope, and the erupting prominence/filament, respectively \citep{2006ApJ...641..590G,2011ApJ...732L..25C,2011LRSP....8....1C}.

To study the dynamics of CMEs, one has to measure their velocity. Usually, people adopt a straightforward method, that is, first determining the height of the leading front above the solar surface, and then linearly fitting the height-time data to derive the average velocity \citep {2004JGRA..109.7105Y}. The statistical results show that the average velocities of CMEs vary from 300 km s$^{-1}$ (in solar minimum years) to 500 km s$^{-1}$ (in solar maximum years). By using the piecewise numerical derivative method, \citet{2001ApJ...559..452Z,2004ApJ...604..420Z} found that the evolution of CMEs involves three different phases: a slow rise phase, an impulsive acceleration phase, and a propagation phase. Nevertheless, the velocity that is measured directly as above is usually subject to a projection effect, except for the ones that occur at the solar limb. In order to remove the projection effect, many advanced methods have been proposed to derive the real velocity of CMEs that actually propagate in three dimensions (3D), such as the geometric triangulation techniques \citep{2010ApJ...710L..82L,2010ApJ...722.1762L} and the Graduated Cylindrical Shell (GCS) model \citep{2006ApJ...652..763T,2009SoPh..256..111T,2011ApJS..194...33T}. Both of the two methods need observations of CMEs in at least two perspectives.

The eruption of CMEs is mostly associated with flares, which appear as a rapid increase of electromagnetic emission at nearly all the wavelengths. Based on the characteristics of the variation of \textit{GOES} soft X-ray (SXR) flux, the evolution of flares is usually divided into three distinct phases, including the pre-flare phase, rise phase, and decay phase. \citet{2006ApJ...649.1100Z} found a close relationship between the three phases of flares and the corresponding three phases of CMEs mentioned above. Moreover, \citet{2008ApJ...673L..95T,2010ApJ...712.1410T} found that the hard X-ray (HXR) emission of a flare is closely associated with the CME acceleration. These results show that the dynamics of CMEs and the energy release of flares are intrinsically coupled to each other.

Besides flares, EUV waves are also often accompanied with CMEs. Usually, EUV waves appear as globally propagating circular fronts when seen on the solar disk \citep[e.g.,][]{1998GeoRL..25.2465T,1999ApJ...517L.151T,2009ApJ...698L.112C,2010ApJ...723L..53L} or outward moving loop-like fronts ahead of the CME fronts as seen at the solar limb \citep[e.g.,][]{2008ApJ...681L.113V,2009ApJ...700L.182P,2010ApJ...716L..57V,2011ApJ...732L..20C,2012ApJ...745L...5C}. After a hot debate of nearly twenty years, most people now recognize that the EUV waves are actually coupled with the CME fronts initially but later on will separate from the CME fronts and then propagate freely as MHD fast waves \citep{2012ApJ...745L...5C,2012ApJ...750..134D}. Moreover, the EUV waves may also evolve into shock waves when their velocities reach a certain value, to say larger than the Alfven speed of the background, which can be evidenced by the appearance of type II radio bursts in some events \citep{1950AuSRA...3..387W,1954AuJPh...7..439W,1985srph.book..333N}.

Although a large number of observations have been devoted to CMEs in the past decades, less attention has been paid to the formation and early evolution of CMEs \citep{2006SSRv..123..251F}. The main reason is that most of the previous instruments like white-light coronagraphs only observe the outer corona \citep{2012LRSP....9....3W}. Nevertheless, the formation and early evolution of CMEs and their associated shock waves take place in the inner corona. This situation has been changed since the launch of the \textit{Solar Dynamics Observatory} \citep[\textit{SDO};][]{2012SoPh..275....3P}. The Atmospheric Imaging Assembly (AIA; \citealp{2012SoPh..275...17L}) on board \textit{SDO} is capable of imaging the inner corona with a field of view (FOV) of 1.3$R_{sun}$. AIA also has a high cadence of 12 s, a high spatial resolution of 1.2{\arcsec}, and a high sensitivity. Thus, it is possible for us to study the formation and early evolution of CMEs and the associated shock waves in detail. Moreover, the six EUV passbands of AIA cover a wide temperature ranging from 0.6 to 20 MK \citep{2010A&A...521A..21O,2012SoPh..275...17L}, which enables us to study the thermal properties of different structural components of CMEs and shock waves.

In this paper, we perform a detailed investigation on the early dynamics of a well observed limb CME that occurred on 2014 January 8. We find that the formation of the CME is a result of the expansion of a bubble-like structure. The CME also causes an EUV wave that further steepens into a shock wave and produces a type II radio burst. In Section \ref{sec:2}, we present observations and results, which are followed by a summary and discussions in Section \ref{sec:3}.

\section{OBSERVATIONS AND RESULTS}
\label{sec:2}
\begin{figure}[!ht]
\centering
\includegraphics[scale=0.9]{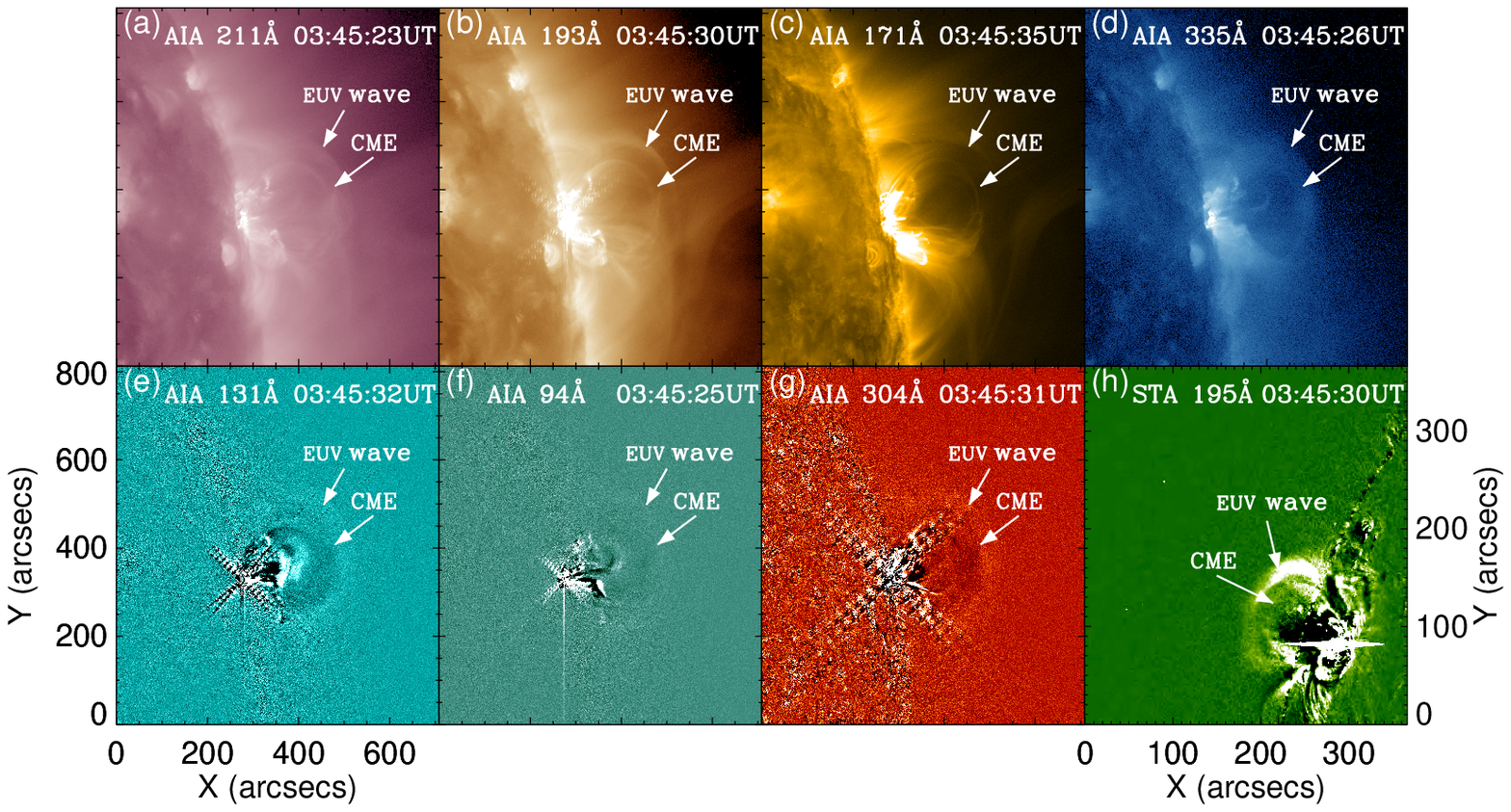}
\caption{(a)--(d) AIA 211 {\AA}, 193 {\AA}, 171 {\AA}, and 335 {\AA} images showing the structures of the CME and the EUV wave. (e)--(g) AIA 131 {\AA}, 94 {\AA}, and 304 {\AA} running difference images. (h) \textit{STEREO}-A 195 {\AA} running difference image showing the CME and the EUV wave from the second perspective.}
\label{fig1}
\end{figure}

On 2014 January 8, there occurred an M3.6 class flare in the active region (AR) NOAA 11947 (N11W91), which started at 03:39 UT and peaked at 03:47 UT. The associated CME firstly appeared as a bubble-like structure in the AIA FOV at $\sim$03:43 UT. About one minute later ($\sim$03:44 UT), an EUV wave firstly came into existence ahead of the CME with a height of 0.1$R_{sun}$ above the solar surface. It is interesting that both the bubble-like structure of the CME and the EUV wave are clearly visible at all of the AIA EUV passbands, as well as at the 195 {\AA} passband of the EUVI on board the \textit{Solar Terrestrial Relations Observatory-A} \citep[\textit{STEREO-A};][]{2008SSRv..136....5K}, which has a separation angle of 150 degree from the Sun-Earth line. Figure \ref{fig1} shows the observations at $\sim$03:45 UT, including seven AIA passbands (211 {\AA}, 193 {\AA}, 171 {\AA}, 335 {\AA}, 131 {\AA}, 94 {\AA}, and 304 {\AA}) and one EUVI (195 {\AA}) passband.

The initial formation of the CME seems to originate from the expansion of coronal loops. Firstly, the loops above the AR began to rise slowly since $\sim$03:43 UT, along with the AR brightening. Then, a bubble-like structure was gradually formed, as best seen at AIA 171 {\AA}, 193 {\AA}, and 211 {\AA} passbands. Subsequently, the structure rose up and expanded rapidly to form the CME. However, about four minutes later, the CME front gradually became vague, which is mainly due to the significant reduction in emission caused by the expansion. As seen at AIA 193 {\AA} and 211 {\AA} passbands, the EUV wave appeared in front of the CME since $\sim$03:44 UT and remained to exist until it left the AIA FOV. In addition, a type II radio burst was observed to start at $\sim$03:47 UT, indicating that the EUV wave had evolved into a shock wave since then. This type II radio burst was well observed by the Learmonth solar radio spectrograph (LSRS) and Siberian Solor Radio Telescope (SSRT).

\subsection{Kinematics of the CME and the EUV Wave}
\label{sec:2.1}
\begin{figure}[!ht]
\centering
\includegraphics[scale=1.2]{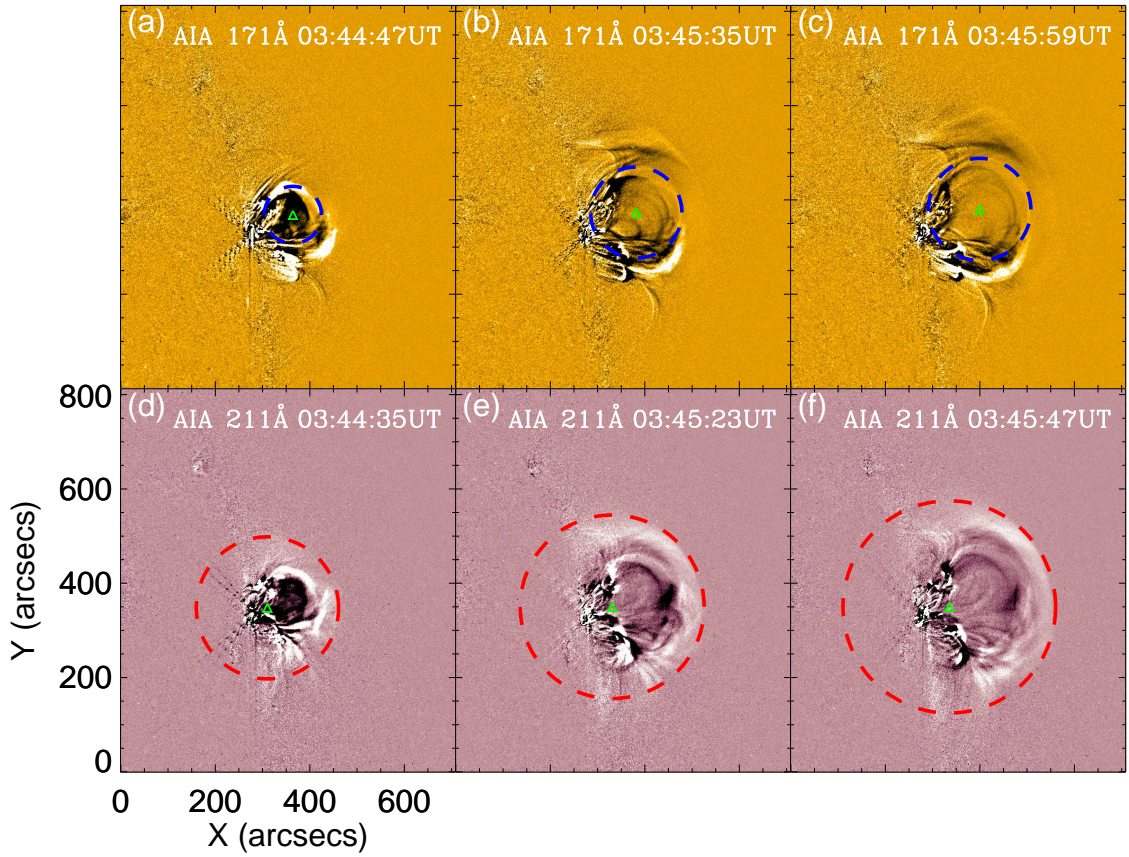}
\caption{AIA 171 {\AA} (top row) and AIA 211 {\AA} (bottom row) running difference images showing the evolution of the circular bubble-like structure of the CME and the semicircular front of the EUV wave, respectively. The blue (red) circles display the fitting to the CME (EUV wave) front. The triangle in each panel marks the circle center. }

\label{fig2}
\end{figure}

AIA high-cadence and multi-wavelength observations enable us to study in detail the dynamics of the CME and the EUV wave. As the CME front and the wave front are nearly circular-shaped in the early phase seen from AIA images, we use a series of circles to fit their morphologies, by varying the centers and sizes of the circles. The circle top is then regarded as the height of the CME front or the wave front. The AIA 171 {\AA} and 211 {\AA} images are chosen for performing such a fitting. Figure \ref{fig2} shows the fitting results, in which one can see that the CME leading front (top row) is well represented by the top parts of the circles in blue and the wave front (bottom row) is well represented by the top parts of the circles in red. Note that, the brightest structure located on the south of the CME refers to some ambient loops that are not associated with the bubble-like structure of the CME.

\begin{figure}[!ht]
\centering
\includegraphics[scale=2]{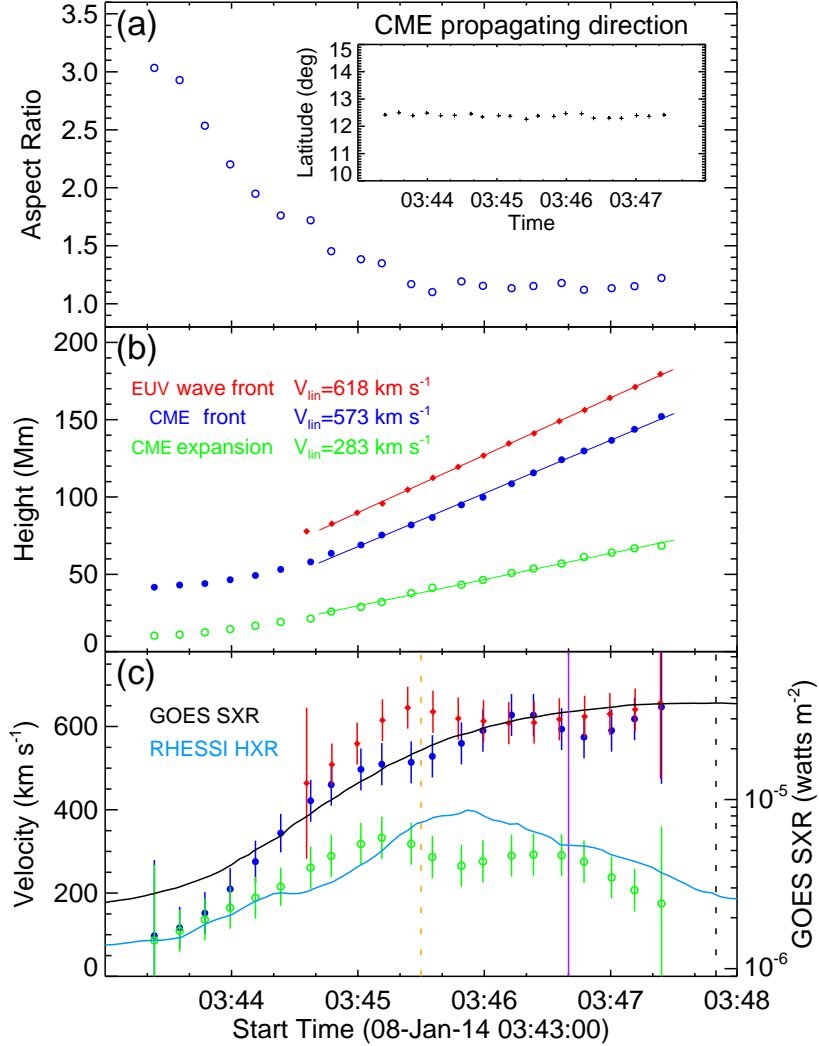}
\caption{Kinematics of the CME and the EUV wave. (a) Temporal evolution of the aspect ratio of the CME. Insert is the variation of the CME propagating direction in latitude. (b) Temporal evolution of the height of the EUV wave front (red), the CME front (blue), and the CME size (green). (c) Temporal evolution of the velocities of the EUV wave front, the CME front, and the CME expansion. At the same time, the velocity difference between the CME front and the CME expansion means its centroid velocity. The associated \textit{GOES} 1--8 {\AA} soft X-ray flux (black) and \textit{RHESSI} 25--50 keV hard X-ray flux (cyan) are also plotted. The three vertical lines mark the start of the type II radio burst (purple), the SXR flux peak time (black dashed), and the time when the CME begins to interact with the inverse Y-shaped structure (orange dashed), respectively.  }
\label{fig3}
\end{figure}

The kinematics of the CME and the EUV wave are shown in Figure \ref{fig3}. One can see that the propagation of the CME is almost radial (inserted panel in Figure \ref{fig3}(a)). An important finding is that the aspect ratio of the CME (defined as the ratio of the fitted circle center height of the bubble-like structure to its size) quickly decreases at the beginning and then keeps a constant after two minutes. It means that the CME first undergoes a fast but short lived lateral over-expansion (i.e., the velocity of the CME expansion is faster than the radial velocity of its centroid) during its initial formation, and then propagates with a self-similar expansion (i.e., the velocity of the expansion and the radial velocity of its centroid are comparable) until it leaves the FOV of AIA.

The temporal evolutions of the height and size of the CME are shown in Figure \ref{fig3}(b), together with the temporal evolution of the height of the EUV wave front. We estimate the uncertainties in height measurement to be two AIA pixels, which are not shown in the figure because the error bars are smaller than the symbol size. It can be seen that the EUV wave gradually separates from the CME front \citep[also see,][]{2012ApJ...745L...5C}. By applying the linear fitting to the height-time data, the average velocities of the EUV wave front and the CME leading front are estimated to be $\sim$618 km s$^{-1}$ and $\sim$573 km s$^{-1}$, respectively. Moreover, the EUV wave becomes visible about one minute after the CME bubble-like structure appears and exists until it leaves the FOV of AIA.

\begin{figure}[!ht]
\centering
\includegraphics[scale=1.3]{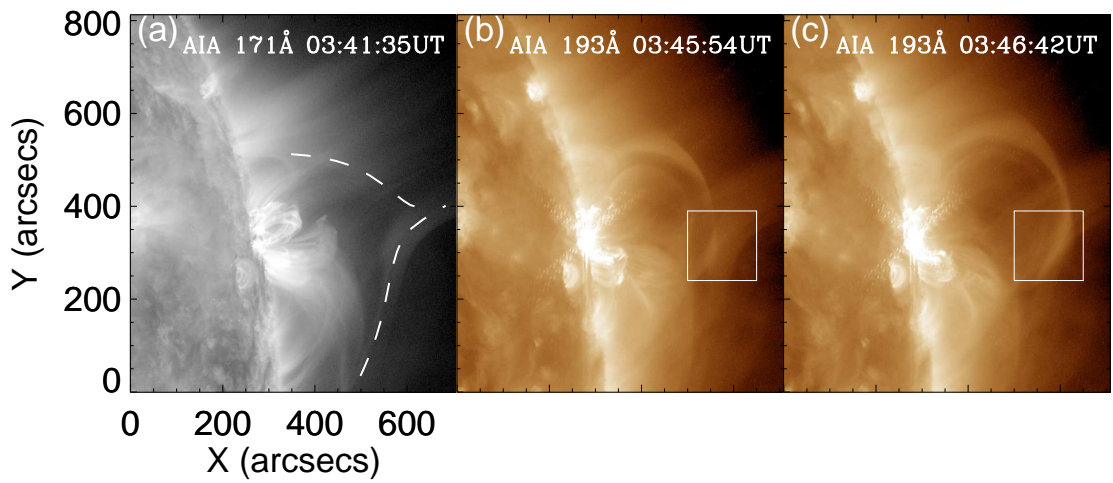}
\caption{(a) AIA 171 {\AA} image showing the preexisting inverse Y-shaped structure. (b)--(c) AIA 193 {\AA} images showing the interaction of the CME with the inverse Y-shaped structure that is highlighted in the small boxes.}
\label{fig4}
\end{figure}

Based on the height-time data, we can also calculate the time-dependent velocity, which is displayed in Figure \ref{fig3}(c). The associated \textit{GOES} 1--8 {\AA} SXR flux (black) and \textit{RHESSI} 25--50 keV hard X-ray (HXR) flux (cyan) are also plotted. At the beginning, with the SXR flux increasing, both the velocity of the CME leading front and the CME expansion velocity grow rapidly. Here note that, the velocity difference between them at the same time means the centroid velocity of the CME. Also, the similarity between them implies that the CME expansion dominates over the radial propagation of the CME in the very early phase. About one minute later, the CME has been accelerated to a relatively high velocity of $\sim$420 km s$^{-1}$. At the same time, the EUV wave becomes visible and also experiences an acceleration process, which is mainly driven by the upward movement of the CME. At about 03:45:30 UT (orange dashed), both the CME and the EUV wave start to slow down in motion. This may be caused by the inverse Y-shaped structure, as shown in Figure \ref{fig4}(a), which interacts with the CME and impedes the continuous acceleration of the CME and the EUV wave (Figure \ref{fig4}(b) and \ref{fig4}(c)).

At $\sim$03:47 UT (purple), a type II radio burst is observed, which suggests that the EUV wave has evolved into a shock wave. At this moment, the velocity of the CME is almost the same as that of the shock wave ($\sim$600 km s$^{-1}$). After that, the shock wave and the CME propagate with a relatively constant velocity. The \textit{GOES} SXR flux reached its peak (black dashed) at about 03:47:50 UT, about two minutes later than the time of the peak velocities of the CME and the shock wave.
Overall, the temporal evolution of the velocities of the CME and the EUV wave front coincides with the variation of the flare SXR flux, implying that the dynamics of the CME and the EUV wave are powered by the same physical process as the flare emission.

\subsection{Shock Wave-produced Type II Radio Burst}
\label{sec:2.2}
\begin{figure}[!ht]
\centering
\includegraphics[scale=1.6]{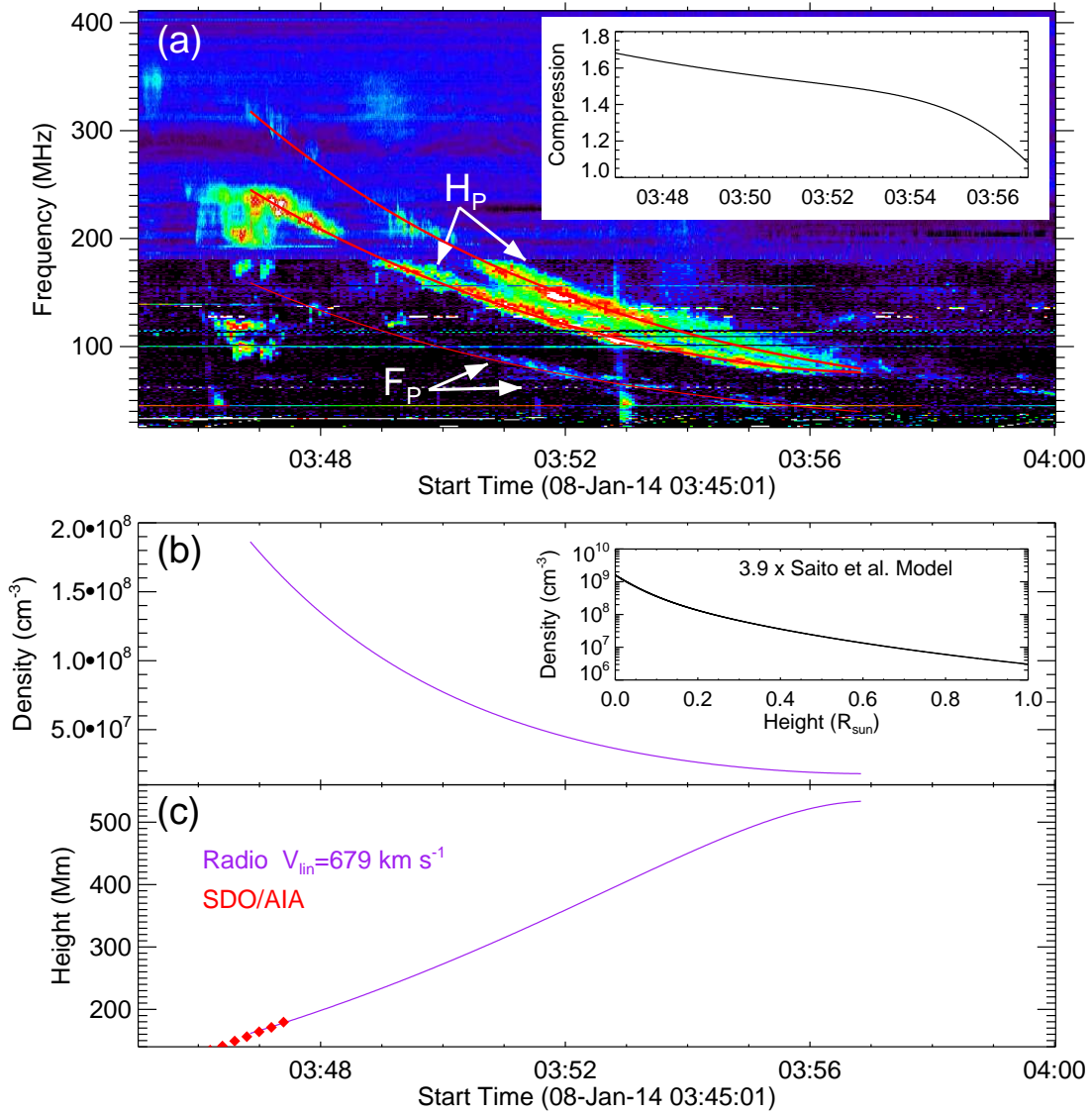}
\caption{(a) Radio dynamic spectrum observed by LSRS (25--180 MHz) and SSRT (180--410 MHz) with a clear band splitting in the harmonic frequencies (the upper two red curves), from which the compression ratio is derived shown in the inserted panel. (b) Temporal variation of the density of the radio burst source region derived from observations (purple). Also shown is the density model in the inserted panel. (c) Temporal evolution of the corresponding height of the source region derived from the radio data and the density model (purple), as well as the height of the shock wave directly measured from AIA images (red).}
\label{fig5}
\end{figure}

We further study the type II radio burst that is generally believed to be produced by the CME-driven shock wave \citep[e.g.,][]{2001JGR...10625279R,2006ApJ...649.1110L,2007SoPh..245..391O,2008SoPh..253..305M,2008SoPh..253..215V}. Although this burst is observed by several radio spectrographs, we only choose the data from LSRS and SSRT based on the clarity of their dynamic spectra. The data from LSRS at low frequencies (25--180 MHz) basically cover the whole radio burst. The data from SSRT at high frequencies (180--410 MHz) provides the information of the early stage, which is used for determining the onset of the radio burst. From Figure \ref{fig5}(a), we can easily identify two branches of the harmonic frequency (2$f_{p}$) that represent the band splitting and one branch of the fundamental frequency that displays the local plasma frequency ($f_{p}$). They first appear at $\sim$03:47 UT and then drift toward the lower frequencies, which are fitted well by double exponential functions (red curves). For the band splitting of the harmonic frequency, we calculate the compression ratio as follows:
\begin{equation}
X = \frac{n_{2}}{n_{1}} = (\frac{f_{2}}{f_{1}})^{2},
\end{equation}
where the subscripts 1 and 2 represent the upstream and downstream of the shock wave, respectively. There often exists a jump of electron density when crossing the shock wave \citep{2002A&A...396..673V}. The density ratio is found to be $\sim$1.68 in this event at the onset time (inserted panel in Figure \ref{fig5}(a)). After that, the density ratio gradually decreases. Furthermore, taking advantage of the relationship between plasma frequency and density,
\begin{equation}
f_{p} = 8.98 \times  10^{3} \sqrt{n},
\end{equation}
we can obtain the plasma density of the shock front (purple) as shown in Figure \ref{fig5}(b). For this event, we adopt the density model proposed by \citet{1977SoPh...55..121S} multiplied by a modification factor of 3.9 (inserted panel in Figure \ref{fig5}(b)). The modification factor is determined by equating the heights of the shock wave directly measured from the AIA images (red) and derived from the dynamic spectrum and density model (purple) \citep[also see,][]{2015ApJ...804...88S}. Note that, there are no radio-heliograph observations at that moment so that we cannot determine the radio source height more accurately \citep{2012A&A...547A...6Z,2014ApJ...795...68Z}.

The derive height of the shock wave (purple) is displayed in Figure \ref{fig5}(c), with a linear velocity of $\sim$679 km s$^{-1}$. Differently from the limited FOV of AIA EUV images, the radio observations can track the shock wave up to the height of $\sim$0.8$R_{sun}$. After 03:57 UT, it is unable to clearly identify the signal caused by the shock wave from the radio data any more. It indicates that the type II radio burst is gradually disappearing, which is also proved by the compression ratio of being close to 1 at that time.

\subsection{Thermal Properties of the CME and the EUV Wave}
\label{sec:2.3}
\begin{figure}[!ht]
\centering
\includegraphics[scale=0.85]{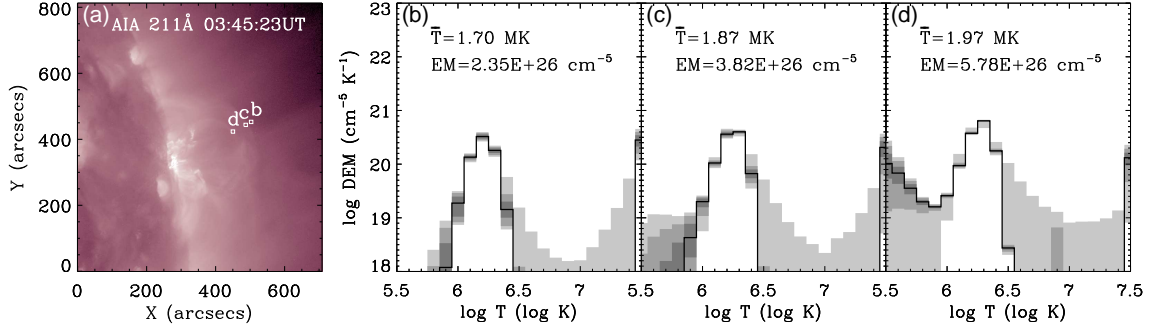}
\caption{(a) AIA 211 {\AA} image showing the CME leading front, the EUV wave front, and the background that are represented by three small boxes, respectively. (b)--(d) DEMs of the three different structures. The best fitted DEMs are plotted in solid curves. The three gray shades (from dark to light) represent the regions that contain 68{\%}, 95{\%}, and 99.7{\%} of the MC simulations, respectively. We use the temperature range of 5.9 $\leq$ log$T \leq$ 6.6 to calculate the DEM-weighted average temperature ($\bar{T}$) and total emission measure ($EM$).}
\label{fig6}
\end{figure}

AIA multi-wavelength observations make it possible to investigate the thermal properties of different structures of CMEs by differential emission measure (DEM) analysis \citep{2004IAUS..223..321W,2012ApJ...761...62C,2013A&A...553A..10H}.
We then perform a DEM analysis of the CME and plot the results at 03:45:23 UT in Figure \ref{fig6} as an example. We select three small boxes (b,c and d in Figure \ref{fig6}(a)) along the direction of CME propagation to characterize the background, the wave front, and the CME front, respectively.
The DEM results of the three features are shown in Figure \ref{fig6}(b)--(d). The black lines refer to the best fitted (i.e., the most possible) DEM solutions from 10,000 Monte Carlo (MC) simulations. The three gray shades (from dark to light) represent the regions that contain 68{\%}, 95{\%}, and 99.7{\%} of the MC simulations, respectively. All of the three regions show relatively narrow temperature distributions. One can see that the emission in these regions mainly comes from the plasma with a temperature of $\sim$2 MK.

In order to quantitatively study the thermal properties of the plasma, we introduce two physical quantities: DEM-weighted average temperature, defined by $\bar{T} =\int { DEM(T){\times}T dT } / \int { DEM(T) dT }$ and total emission measure, defined by $ EM=\int {DEM(T) dT} $. A temperature range from $\sim$0.8 MK to $\sim$4 MK (or $5.9 \leq log T \leq 6.6$) is used for the $EM$ integration because the DEM solutions are well constrained within this range. One can see that the average temperatures decrease with height of the three different components: $\sim$1.97 MK for the CME front, $\sim$1.87 MK for the wave front, and $\sim$1.7 MK for the background. In addition to the slight change of the temperature, the emission measures vary a lot in the different components. The results above indicate that the EUV wave and the CME leading front can slightly heat and compress the plasma. The compression ratio is estimated as follows:
\begin{equation}
X = \frac{n_{2}}{n_{1}} = \frac{\sqrt{EM_{2}/l}}{\sqrt{EM_{1}/l}}=\sqrt{ \frac{EM_{2}}{EM_{1}}},
\end{equation}
where $l$ is the depth along the line of sight. Assuming that the depth does not change \citep{2011ApJ...733L..25K,2014SoPh..289.2123K}, the value of $X$ is calculated to be $\sim$1.27 ($\sim$1.57) for the top of the wave (CME) front. Note that, for the CME front, the compression ratio may represent an upper limit because its real background emission is hard to determine and we thus use that of the wave front instead (the region b in Figure \ref{fig6}(a)), which is slightly smaller.

\begin{figure}[!ht]
\centering
\includegraphics[scale=1.15]{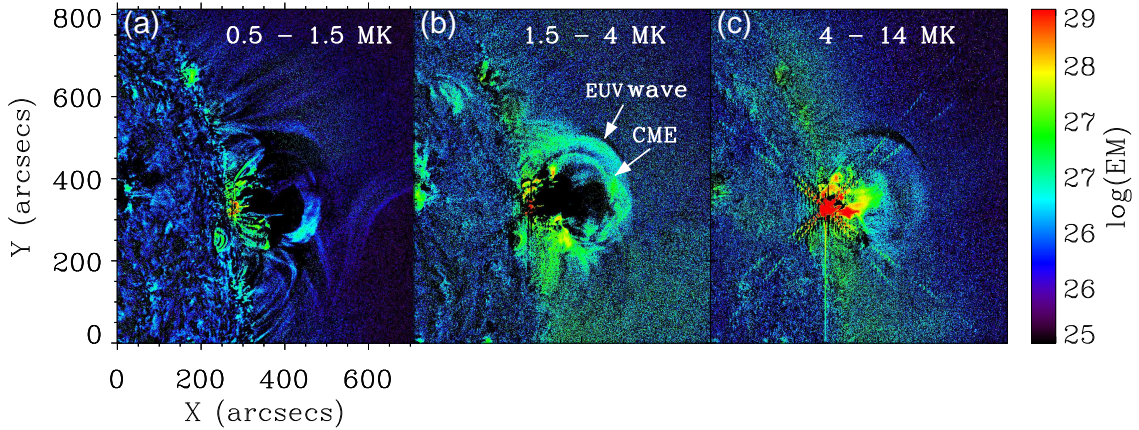}
\caption{EM maps in three different temperature ranges showing the thermal proprieties of the background (a), the CME structures, the EUV wave (b), the flare region, and the CME core region (c). For clarity, the EM maps have been subtracted by those before the beginning of the flare.}
\label{fig7}
\end{figure}

We also construct the EM maps of the CME at different temperature ranges. In order to see different structures more clearly, we subtract the EM maps by those at 03:19:25 UT (before the flare beginning). The results are shown in Figure \ref{fig7}, from which one can clearly see that the emission measure in the CME leading front and the EUV wave front is mostly contributed by plasma in the temperature range of 1.5--4 MK. However, the emission measure in the flare region and the CME core region is mainly from hotter plasma with temperatures of 4--14 MK. As for the hot plasma in the CME core region, we suspect that it most likely originates from the hot flux rope structure that has been confirmed by \cite{2015ApJ...808..117N}.

\section{SUMMARY AND DISCUSSIONS}
\label{sec:3}
In this paper, we study the formation and early evolution of a limb CME and its associated shock wave on 2014 January 8. The event is very rare since the CME leading front and the EUV wave front are simultaneously visible at all of the AIA EUV passbands. Through kinematical analyses, we find that the CME early formation is mainly from the expansion of a bubble-like structure that undergoes two phases: a strong but short-lived lateral over-expansion phase and a self-similar expansion phase. Moreover, the expansion of the CME also drives an EUV wave that further steepens into a shock wave and produces a type II radio burst.

The formation and early evolution of the current CME is comparable with that of the 2010 June 13 CME analyzed by \citet{2010ApJ...724L.188P}. They found that the CME formation undergoes three phases: a slow self-similar expansion phase, a fast but short-lived over-expansion phase, and a second self-similar expansion phase. For the current event, the slow self-similar expansion phase does not appear. Moreover, they found that the over-expansion phase is more related to the flare decay phase rather than the rise phase, thus suggesting that the ideal MHD effect may play a role in driving the expansion of the CME. While in our case, the temporal evolution of the CME expansion coincides very well with the variation of the HXR flux of the flare, thus favoring the role of magnetic reconnection as the main driving mechanism \citep[also see,][]{2010A&A...522A.100P}.

We also study the formation of the CME-associated shock wave. It is found that, during the initial expansion process of the CME, there appears a second quasi-circular bright front, i.e., the EUV wave, which can clearly be distinguished from the CME front above $\sim$0.1$R_{sun}$. Subsequently, the EUV wave moves outward and is accelerated by the eruption of the CME \citep[also see,][]{2014ApJ...797...37L}. Then, it becomes a shock wave when reaching a relatively high velocity of $\sim$600 km s$^{-1}$, which is evidenced by the appearance of the type II radio burst. The formation height of the shock wave is estimated to be $\sim$0.2$R_{sun}$, which is smaller than the height of $\sim$0.3$R_{sun}$ for the 2010 June 13 event \citep{2011ApJ...738..160M} and also the average height of 0.5$R_{sun}$ for STEREO CMEs \citep{2009SoPh..259..227G}. Through the band splitting of the radio spectrum, the compression ratio of the shock wave is estimated to be $\sim$1.68 at $\sim$03:47 UT (starting of the type II burst), which is obviously larger than the compression ratio of $\sim$1.27 at $\sim$03:45 UT near the top of the EUV wave front. It indicates that, although the EUV wave appears as early as 03:45 UT, only if it has been accelerated to a certain velocity, e.g., $\sim$600 km s$^{-1}$, it can evolve into a shock wave that in turn produces a type II radio burst.

We further construct the EM maps of the CME and the EUV wave, which show that the low temperature plasma is dominant in the CME front and the wave front, while the high temperature plasma is dominant in the flare region and the CME core. Such results are consistent with the results of \cite{2012ApJ...761...62C}. Moreover, it is revealed that the CME actually contains a hot structure, as can be seen in the EM map at the temperature range of 4--14 MK. Such a hot structure may correspond to a hot flux rope that has been reported in many other events \citep{2012NatCo...3E.747Z,2015ApJ...808..117N}. However, in the current case, the hot structure just appears after the beginning of the flare. We conjecture that the preexisting flux rope is still cool and thus invisible before the flare but it quickly gets heated after the flare begins.

\acknowledgements
We thank the anonymous referee for his/her constructive comments that improved the manuscript significantly. We are also grateful to Pengfei Chen, Yu Dai, Jie Hong and Kai Yang for their helpful discussions. \textit{SDO} is a mission of NASA's Living With a Star Program, \textit{STEREO} is the third mission in NASA's Solar Terrestrial Probes program. This work is supported by NSFC under grants 11303016 and 11373023, and by NKBRSF under grant 2014CB744203.

\end{document}